\documentclass[prl,twocolumn,showpacs]{revtex4}
\usepackage{graphicx}
\usepackage{psfig}
\renewcommand{\nocite}[1]{}
\begin{document}
\title{Signature of directed chaos in the conductance of a nanowire} 
\author{Manamohan Prusty} 
\author{Holger Schanz} 
\email{holger@chaos.gwdg.de}
\affiliation{Max-Planck-Institut f\"ur Dynamik und Selbstorganisation, 
und Fakult{\"a}t f{\"u}r Physik, Universit{\"a}t G{\"o}ttingen,
Bunsenstra{\ss}e 10, D-37073 G\"ottingen, Germany}
\date{\today}
\begin{abstract}
  We study the conductance of chaotic or disordered wires in a situation where
  equilibrium transport decomposes into biased diffusion and a counter-moving
  regular current. A possible realization is a semiconductor nanostructure
  with transversal magnetic field and suitably patterned surfaces. We find a
  non-trivial dependence of the conductance on the wire length. It differs
  qualitatively from Ohm's law by the existence of a characteristic length
  scale and a finite saturation value.
\end{abstract}
\maketitle
\def\sr{R}
\def\rc{r_{c}}
\def\reg{_{\rm reg}}
\def\ch{_{\rm ch}}
\def\uc{_{\rm uc}}
\def\ve{v_{\rm F}}
\def\t{t}
\def\tp{t'}
\def\tpr{\tp\reg}
\def\tpc{\tp\ch}
\def\delx{\Delta x}
\def\dell{\Delta l}
According to Ohm's law, the resistance of a wire is proportional to its
length. This is a straightforward consequence of the diffusive motion of
electrons in the disordered potential of a normal material. However, unlike
the time when Ohm arrived at his fundamental observation, conductors can be
tailor-made today with almost complete control over the microscopic structure.
Therefore it is important to understand the consequences of non-diffusive
electron dynamics on the electronic conductance or other transport properties.
This question has been studied in much detail for semiconductor
nanostructures in which the motion of electrons is ballistic rather than
diffusive
\nocite{Imry}\nocite{M+92,H+98a}\nocite{P+01}\nocite{W+91,W+93}\cite{Imry,M+92,P+01,W+91}.
In such systems disorder is negligible and consequently all transport
properties are determined by the shape of the sample, as in a billiard model.
For example, in the ideal case of a perfectly clean nanowire with parallel
walls the resistance should be zero independent of the length, and indeed this
remarkable prediction has been confirmed experimentally \cite{P+01}.  Beside
ballistic systems, also the effects of anomalous diffusion on the electronic
or thermal conduction properties have attracted a lot of attention
\nocite{FGK94}\nocite{LFL98}\nocite{LW03,DKU03}\cite{FGK94,LFL98,LW03}.

In the present paper we study the electronic conductance of a wire in the case
of a different and very profound modification of the microscopic dynamics.  We
consider systems where {\em directed chaos} leads to biased diffusion in the
absence of any potential gradient.  Directed chaos means that the
time-averaged velocity of chaotic trajectories is non-zero due to broken
time-reversal symmetry and due to the specific phase space structure. This
effect may occur in various types of systems including, e.g., cold atoms in
suitably pulsed optical potentials or chains of electronic billiards in a
transversal magnetic field. It has been investigated both, theoretically and
experimentally, in a number of recent publications
\nocite{FYZ00}\nocite{S+01,SDK05}\nocite{M+02,J+05}
\cite{FYZ00,S+01,M+02,AD03,SP05}. The interest is due to some intriguing and
potentially very useful properties. For example, quantities like velocity
average, velocity dispersion or scattering delay times are intrinsically
dependent on the transport direction. However, all previous studies focused on
the ratchet-like directed transport in effectively infinite periodic systems
with directed chaos. In contrast, we address here for the first time the
typical electronic setup of a finite sample which is coupled to two electron
reservoirs. We show that directed chaos has profound consequences also in this
context where the transport velocity is not directly measurable; instead the
conductance becomes the most basic and most relevant quantity. Due to biased
diffusion, a new length scale $\lambda\ch$ appears and rules the asymptotic
decay of the conductance with the sample length, see Eq.~(\ref{expdec}) below.
Moreover, chaotic trajectories can propagate through samples of arbitrary
length and consequently the conductance approaches a non-zero constant
given in Eq.~(\ref{tinf}). Note that it is not possible to reduce the
description of directed chaos to a diffusion equation with 
bias.  In our explicit result for the conductance,
Eq.~(\ref{tottrans}), we must account also for the detailed structure of the
underlying mixed phase space.

\begin{figure}[htb]
 \centerline{\psfig{figure=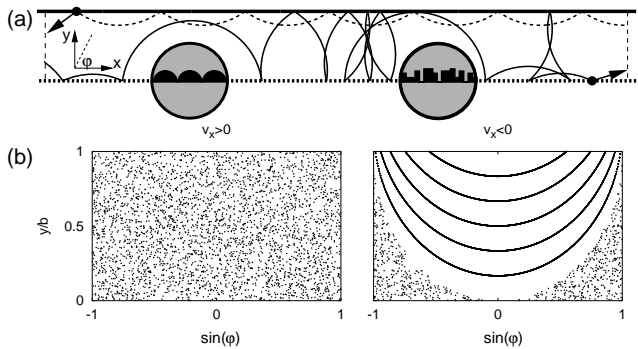,width=85mm}}
 \caption{\label{fig1} (a) A prototypical example for directed chaos is a
   2DEG in a perpendicular magnetic field which is confined by two parallel
   walls with different surfaces. The upper wall reflects specularly while at
   the lower wall the reflection angle is essentially random.  Possible
   physical realizations of this randomness are magnified in the insets (see
   text). Typical trajectories are either regular (dashed line) or random (full
   line) and transport in opposite directions. In (b)
   the transversal Poincar\'e sections for $v_{x}>0$ and $v_{x}<0$ are
   displayed.}
\end{figure}

To be specific, we investigate the prototypical model first introduced in
\cite{SP05}. We consider a two-dimensional electron gas (2DEG) confined to a
quasi one-dimensional channel (Fig.~\ref{fig1}a). One wall of the channel is
straight. Electrons are specularly reflected, but no back scattering occurs in
the transport direction. The other wall has a rough surface causing strong and
essentially random scattering. The details of this roughness are not crucial
(see below).  Directed chaos is induced by a perpendicular magnetic field
which breaks time-reversal invariance. We stress that a realization of our
model does not require more than a novel combination of elements which are all
well understood and experimentally accessible in the context of a mesoscopic
2DEG, namely transversal magnetic fields of moderate strength, negligible bulk
scattering and surfaces which are either disordered or manufactured with a
precisely defined geometry \cite{Imry,M+92,P+01,W+91,FGK94,LFL98}.

For our purpose it is sufficient to treat the electrons as independent
classical particles, i.e., we assume a semiclassical regime with many
transversal modes in the channel, $N\sim m^{*}\!\ve b/h\gg1$. We use
dimensionless units in which channel width and Fermi velocity are unity,
$b=\ve=1$.  The strength of the magnetic field is para\-mete\-rized by the
cyclotron radius $\rc=m^{*}\!\ve/eB$. Besides the length of the sample this is
the only free parameter in our problem. We will assume that the magnetic field
is not too strong, $\rc\ge 1$. In this case there are no pinned orbits in the
bulk of the channel and the phase space contains only two types of electron
trajectories: regular orbits skipping along the clean channel boundary, and
chaotic or random orbits which are reflected from both walls
(Fig.~\ref{fig1}a). The regular orbits are transporting continuously in one
direction, say to the left. On the average, the chaotic orbits are
transporting in the opposite direction, thus compensating for the regular
transport and making the system unbiased as a whole. However, the transport
velocity $\dot x$ of a chaotic electron is fluctuating, in fact the dynamics
is diffusive with a superimposed drift along the channel. The average drift
velocity can be obtained by application of a phase space sum rule \cite{S+01}.
For our model we find for the long-time velocity average of almost all chaotic
trajectories
\begin{eqnarray}\label{vchr}
v\ch&=&{1\over 2}{\theta-\sin\theta\cos\theta\over \pi(1-\cos\theta)-
\sin\theta+\theta\,\cos\theta}
\end{eqnarray}
with $\theta(\rc)=\arccos(1-{1/\rc})$ \cite{SP05}. This result is
independent of the precise modelling of the rough channel surface as long as
the phase space structure of Fig.~\ref{fig1}b is preserved. In \cite{SP05} we
considered the two extreme cases shown in the insets of Fig.~\ref{fig1}a. In
one case the surface is a periodic array of semicircular scatterers with small
radius $\sr\to 0$ \endnote{Finite $\sr$ results in a correction to the first
  term in the denominator of Eq.~(\ref{vchr}), see \cite{SP05}.}, i.e., the
dynamics of the system is deterministic and there is no disorder whatsoever.
In the second case the direction $\varphi$ of the trajectory was randomized
upon every scattering from the rough surface. Specifically it was chosen with
probability density $P(\varphi)={1\over 2}{\sin\varphi}$ from the interval
$[0,\pi]$ such that the invariant measure on the energy shell $d x\,d y\,d
\varphi$ was preserved. A physical realization of this behavior is a
disordered surface with a correlation length that is below the Fermi
wavelength. While this second system is non deterministic when approximated
classically, its transport properties are essentially the same as for the case
of deterministic directed chaos. In particular, in both cases an ensemble of
chaotic trajectories spreads diffusively around the moving center of mass,
$\langle\Delta x^2\rangle=D\ch t$ \cite{SP05}. The precise value of the diffusion
constant depends on the detailed modelling of the rough boundary. Analytical
results for $D\ch$ are available in the case of random scattering \cite{SP05}
and therefore we shall use this version of the model in the numerical
calculations below.

\begin{figure}[htb]
 \centerline{\psfig{figure=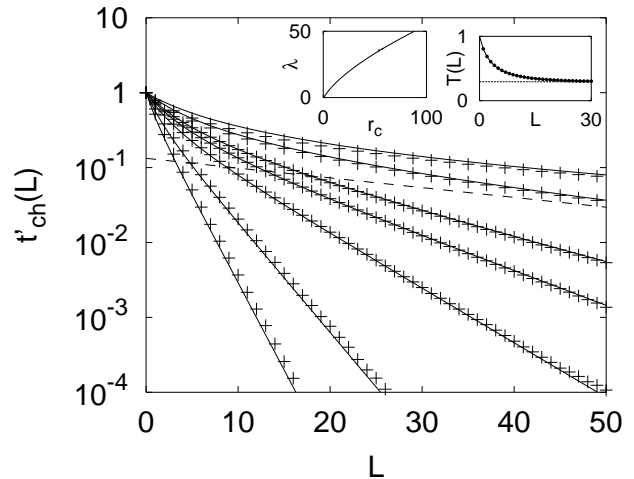,width=85mm}}
 \caption{\label{fig2} The chaotic transmission (+) is compared
   to Eq.~(\ref{tpc}) for $\rc=200, 50, 15, 10, 6, 3, 2$ (top-bottom). Each
   data point represents $10^6$ trajectories. For $\rc=50$ the asymptotic
   exponential is shown with a dashed line. Left inset: Dependence of the
   Peclet length on the cyclotron radius. Right inset: For $\rc=10$ the total
   transmission (\textbullet) is compared to Eq.~(\ref{tottrans}) and to the
   asymptotic constant $\mu\reg=0.294$.  }
\end{figure}

The electronic conductance is obtained within the framework of the
Landauer-B{\"u}ttiker formula \cite{Imry}, $G(L)=(2e^2/h)\,T(L)$. We approximate
the transmission semiclassically, $T\sim N t(L)$, and discuss quantum corrections at the
end of this letter. $t(L)$ denotes the total classical probability that an
electron is transmitted through a sample of length $L$ if it enters the system
at $x=0$ from the left with random initial conditions
$P(y,\varphi)={1\over 2}{\cos\varphi}$ ($y\in[0,1]$,
$\varphi\in[-{\pi\over 2},+{\pi\over 2}]$). According to our convention (regular
orbits are skipping to the left) these initial conditions are within the
chaotic component of phase space. Similarly, we define the probability $t'(L)$
that an electron is transmitted if it enters at $x=L$ from the right with an
analogous distribution (but $\varphi\in[{\pi\over 2},{3\pi\over 2}]$). The total
transmission probability must be the same for the two distinct transport
directions,
\begin{equation}\label{tpeq}
t(L)=t'(L)\,.
\end{equation}
There are various ways to arrive at this fundamental identity. For example,
one observes that $\t\ne\tp$ would result in the accumulation of particles in
one of the reservoirs even when the system is in thermal equilibrium.
Alternatively, a microscopic derivation can be based on the fact that the
scattering map of Hamiltonian systems is area preserving.  Although
Eq.~(\ref{tpeq}) is not specific for directed chaos, this identity has very
interesting consequences in the present context. While $t(L)$ is entirely due to
chaotic trajectories whose properties are not immediately accessible, $t'(L)$
can be decomposed into conditional probabilities for regular and chaotic
trajectories,
\begin{eqnarray}\label{sumtrans}
\tp(L)&=&\mu\reg\,\tpr(L)+\mu\ch\,\tpc(L)\,,
\end{eqnarray}
where 
\begin{equation}
\mu_{\rm reg}={1\over 2}\int\reg dy\,d\varphi\,\cos\varphi
={1\over 2}{\theta-\sin\theta\,\cos\theta\over 1-\cos\theta}
\end{equation}
is the the relative area of the regular component in the transversal
Poincar\'e section (Fig.~\ref{fig1}b), and $\mu\ch=1-\mu\reg$.  Due to the
lack of back scattering along the regular trajectories we have $\tpr(L)\equiv
1$.  Moreover, it is clear that $\tpc\to 0$ for $L\to\infty$. This is so
because for trajectories which are moving through arbitrarily long samples
{\em against} the average chaotic flow the time-averaged velocity cannot
converge to $v\ch$. Hence these trajectories must be of measure zero in phase
space.  Taken together, the mentioned facts yield a remarkable result
which is illustrated in the right inset of Fig.~\ref{fig2}: for long systems the
probability that a {\em chaotic} trajectory transmits from $x=0$ to $x=L$ is given
by the relative phase space area occupied by the {\em counter-moving regular}
trajectories,
\begin{equation}\label{tinf}
t(L)\to\mu\reg\qquad(L\to\infty)\,.
\end{equation}
The goal is now to understand quantitatively how $\tpc(L)$ decays from
$\tpc(0)=1$ to zero as the length of the system increases. The results of
numerical simulations for a wide range of values $\rc$ are shown in
Fig.~\ref{fig2}. The data suggest an exponential behavior for long wires. This can
be understood after replacing the microscopic dynamics by the Fokker-Planck
equation (FPE) for biased diffusion. At $x=0$ and $x=L$ we assume absorbing
boundary conditions. The probabilities to reach these exits from a point $0\le x\le
L$ within the sample are then
\begin{eqnarray}\label{exitprob}
p'_{L}(x)&=&{e^{-x/\lambda\ch}-e^{-L/\lambda\ch}\over 1-e^{-L/\lambda\ch}}\,,
\\[2mm]
p_{L}(x)&=&{1-e^{-x/\lambda\ch}\over 1-e^{-L/\lambda\ch}}
\,,
\end{eqnarray}
respectively \cite{Gardiner}. Here
\begin{equation}\label{lambda}
\lambda\ch=D\ch/v\ch
\end{equation}
denotes the Peclet length of the biased diffusion process. 

In order to contribute to the transmission from the right to the left end of a
wire of length $L$, a chaotic particle should first be transmitted through a
segment of length $l<L$ and then, starting from $x=L-l$, be absorbed at $x=0$.
Based on this argument we propose as a recursion relation for the chaotic
transmission
\begin{eqnarray}\label{recrel}
\tpc(L)=\tpc(l)\,p'_{L}(L-l)\,.
\end{eqnarray}
Note that this relation cannot be valid for arbitrary $l$. For example, $l=0$
leads to $\tpc(L)=0$ as for a diffusing particle starting at one of the
absorbing boundaries the probability to reach the opposite end is identically
zero. The reason for this limitation can be understood as follows. Upon
replacing the original dynamical system by a 1D FPE we have discarded all
information about the momentum of the electron. Strictly speaking it is then
impossible even to define a transmission probability since this requires to enter the
system with a given direction. Hence, for Eq.~(\ref{recrel}) to be valid, the
length scale for momentum correlations should be negligible compared to the
distance from the boundaries, $\lambda\ch\ll x$ and $\lambda\ch\ll l$. Under
this assumption we find to leading order
$\tpc(L)=\tpc(l)\,e^{-(L-l)/\lambda\ch}$ with the solution
\begin{equation}\label{expdec}
\tpc(L)=c\,\exp(-L/\lambda\ch)\,.
\end{equation}
The dashed line in Fig.~\ref{fig2} shows this exponential for $\rc=50$ (with
suitably chosen prefactor). Indeed the asymptotic decay of the transmission
probability is reproduced. However, for short systems the behavior is clearly
not exponential. Moreover, even in the asymptotic regime $L\to\infty$ the FPE
approach is not accurate enough to predict the prefactor $c$ of the
exponential decay \endnote{Extrapolation of Eq.~(\ref{expdec}) to $L=0$ yields
  $c=1$. It is also possible to repeat the argumentation leading to
  Eq.~(\ref{recrel}) for $t(L)$. Using Eq.~(\ref{sumtrans}), one finds in this
  way $c=\mu\reg/\mu\ch$. Both results disagree with the correct value
  $c=\mu\reg$.}.  This is no surprise. According to Eq.~(\ref{expdec}), a
different prefactor corresponds to an additive constant in the system length $L$
and this type of error must be expected after discarding the correlation
length of the momentum.

We conclude that a satisfactory theory for $\tpc(L)$ cannot be based on the
FPE alone. As an alternative model let us consider a generalization of the
persistent random walk \nocite{Weiss,Wei02}\cite{Weiss} to a biased walk
(BPRW): A particle moves with velocity $\pm\ve$ and is reflected with
direction-dependent probabilities at obstacles with spacing $L_{1}$. For a
segment of length $L=nL_{1}$ we denote the transmission and reflection
probabilities by $t_{n}=1-r_{n}=t(nL_{1})$ (left to right) and
$t'_{n}=1-r'_{n}=\tpc(nL_{1})$ (right to left). A multiple-scattering
expansion allows to express these probabilities in terms of a single segment.
We find
\begin{eqnarray}\label{tn}
t_{n}&=&{1-r_{1}'/r_{1}\over (t_{1}'/t_{1})^{n}-r_{1}'/r_{1}},
\\ 
\label{tnp}
t'_{n}&=&{1-r_{1}/r_{1}'\over (t_{1}/t_{1}')^{n}-r_{1}/r_{1}'}\,.
\end{eqnarray}
In our case the chaotic transport is biased to the right. Therefore we assume
$t_{1}>t_{1}'$ and find in the limit $n\to\infty$
\begin{equation}
t_{\infty}=1-r_{1}/r_{1}'
\end{equation}
and
\begin{equation}\label{expdectmat}
t_{n}'=t_{\infty}\,({t_{1}/t_{1}'})^{-n}\qquad(n\to\infty)\,.
\end{equation}
However, there is no direct connection between the parameters of the BPRW and
the dynamics of our original model. In order to close this gap we must make
use of the information about the underlying phase space structure,
Eq.~(\ref{tinf}), and the result obtained within the FPE approach,
Eq.~(\ref{expdec}). The former implies $t_{\infty}=\mu\reg$ or equivalently
$r_{1}/r_{1}'=\mu\ch$.  Further the comparison of Eqs.~(\ref{expdec}) and
(\ref{expdectmat}) yields
\begin{equation}
c\,\exp(-nL_{1}/\lambda\ch)=\mu\reg\,(t_{1}/t_{1}')^{-n}\,.
\end{equation}
Now it is easy to read off the correct prefactor of the asymptotic exponential
decay, $c=\mu\reg$, which was used for the dashed line in Fig.~\ref{fig2}. On
the other hand we infer $\exp(L_{1}/\lambda\ch)=t_{1}/t_{1}'$ which is
substituted into Eqs.~(\ref{tn}) and (\ref{tnp}). The final result for the
chaotic transmission probability from right to left is then
\begin{equation}\label{tpc}
\tpc(L)={\mu\reg\over \exp(L/\lambda\ch)-\mu\ch}
\end{equation}
while the total transmission is given by
\begin{equation}\label{tottrans}
t(L)={\mu\reg\over 1-\mu\ch\,\exp(-L/\lambda\ch)}\,.
\end{equation}
This non-trivial prediction, which does not contain any free parameters, is
confirmed numerically in Fig.~\ref{fig2}. The fact that an appropriate
synthesis between FPE and BPRW should be used to reproduce the data was not at
all obvious. Note that in an analogous approach to the unbiased case the PRW
yields Ohm's law in the form $t(L)=(D/\ve)L^{-1}$ $(L\to\infty)$ while the FPE
implies only $t\sim L^{-1}$. In contrast, for the biased case both approaches
contribute complementary information and are in fact {\em mutually
  incompatible} approximations \endnote{For the BPRW we derive explicit
  expressions for transport velocity $v$ and diffusion constant $D$. Within
  the FPE we must then expect a decay of the transmission probability with
  length scale $\lambda=D/v$. In general this disagrees with the value
  $\tilde\lambda=L_{1}/\ln(t_{1}/t_{1}')$ obtained from
  Eq.~(\ref{expdectmat}).  A continuum limit ($L_{1},r_{1},r_{1}'\to 0$ for
  fixed $t_{\infty}$) is not sufficient to remove this discrepancy.}.

Small but systematic deviations are visible in Fig.~\ref{fig2} for both, very
small and very large cyclotron radius. At least partially these deviations can
be attributed to {\em direct trajectories} which escape from the system before
being scattered. Such trajectories are always sensitive to the details of a
given model and therefore not of primary interest here. We restrict the
discussion to some simple examples.  Trajectories which enter at $x=0$ with a
steep angle $-\pi/2\lesssim\varphi$ will reach $x=0$ again after completing a
simple arc. From the geometric condition $\cos\varphi<y/2\rc$ we see that the
corresponding phase space volume vanishes as $\rc^{-1}$. Therefore these
trajectories lead to deviations for small cyclotron radius. For short systems
or for large cyclotron radius and shallow incidence $\varphi\approx 0$ there
can also be directly transmitted trajectories. A rough estimate requires
$L\lesssim\sqrt{8\rc}$ for their existence which is compatible with
Fig.~\ref{fig2} ($L=40$ for $\rc=200$).
 
According to Eq.~(\ref{tottrans}) the conductance of a wire with directed
chaos saturates to a finite value as $L\to\infty$. This claim is in sharp
contrast to quantum localization which leads to vanishing conductance in any
coherent quasi 1D quantum system with uncorrelated disorder. These two
contradicting statements can be reconciled as follows. In the presence of
directed chaos the localization length $\xi$ diverges exponentially with the
number of transversal modes ($\ln\xi\sim N\sim h^{-1}$) \cite{H+02}. Thus,
even for moderate $N$, the localization length will easily exceed the sample
length $L$ or the coherence length of the given material. In this regime we
can safely ignore localization. Other quantum effects like weak localization
or tunneling are not expected to change our results qualitatively although
they may lead to small corrections $\sim N^{-1}$ in the relevant parameters
$\lambda$ and $\mu\reg$. A numerical analysis of such effects will be
attempted elsewhere.

\end{document}